\documentclass[epj,nopacs]{svjour}
\usepackage{graphics}

\begin{document}

\title{\boldmath
Study of dynamics of the process $e^+e^-\to \pi^+\pi^-\pi^0$ in the
energy range 1.15--2.00 GeV} 
\authorrunning{M.~N.~Achasov et al.}
\titlerunning{
Study of dynamics of the process $e^+e^-\to \pi^+\pi^-\pi^0$ in the
energy range 1.15--2.00 GeV} 

\author{{\large The SND Collaboration}\\ \\
M.~N.~Achasov\inst{1,2} \and
A.~Yu.~Barnyakov\inst{1,2} \and
A.~A.~Baykov\inst{1,2} \and
K.~I.~Beloborodov\inst{1,2} \and
A.~V.~Berdyugin\inst{1,2} \and
D.~E.~Berkaev\inst{1,2} \and
A.~G.~Bogdanchikov\inst{1}\and
A.~A.~Botov\inst{1}\and
T.~V.~Dimova\inst{1,2} \and
V.~P.~Druzhinin\inst{1,2} \and
V.~B.~Golubev\inst{1} \and
A.~N.~Kirpotin\inst{1} \and
L.~V.~Kardapoltsev\inst{1,2} \and
A.~S.~Kasaev\inst{1} \and
A.~G.~Kharlamov\inst{1,2} \and
I.~A.~Koop\inst{1,2} \and
A.~A.~Korol\inst{1,2} \and
D.~P.~Kovrizhin\inst{1} \and
A.~S.~Kupich\inst{1}\and
K.~A.~Martin\inst{1}\and
N.~A.~Melnikova\inst{1}\and
N.~Yu.~Muchnoy\inst{1,2} \and
A.~E.~Obrazovsky\inst{1}\and
E.~V.~Pakhtusova\inst{1}\and
K.~V.~Pugachev\inst{1,2} \and
D.~V.~Rabusov\inst{1}\and
Yu.~A.~Rogovsky\inst{1,2} \and
Y.~S.~Savchenko\inst{1,2} \and
A.~I.~Senchenko\inst{1,2} \and
S.~I.~Serednyakov\inst{1,2} \and
D.~N.~Shatilov\inst{1}\and
Yu.~M.~Shatunov\inst{1,2} \and
D.~A.~Shtol\inst{1} \and
D.~B.~Shwartz\inst{1,2} \and
Z.~K.~Silagadze\inst{1,2} \and
I.~K.~Surin\inst{1} \and
M.~V.~Timoshenko\inst{1} \and
Yu.~V.~Usov\inst{1} \and
V.~N.~Zhabin\inst{1} \and
V.~V.~Zhulanov\inst{1,2}}

\institute{Budker Institute of Nuclear Physics, SB RAS, Novosibirsk, 630090,
Russia \and
Novosibirsk State University, Novosibirsk, 630090, Russia
}
\date{}
\abstract{
The dynamics of the process $ e^+e^- \to \pi^+\pi^-\pi^0 $ is studied
in the energy region from 1.15 to 2.00 GeV using data accumulated 
with the SND detector at the VEPP-2000 $e^+e^-$ collider. 
The Dalitz plot distribution and $\pi^+\pi^-$ mass spectrum are analyzed
in a model including the intermediate states $\rho(770)\pi$, $\rho(1450)\pi$, 
and $\omega\pi^0$. As a result, the energy dependences of the 
$\rho(770)\pi$ and $\rho(1450)\pi$ cross sections and the relative phases
between the $\rho(770)\pi$ amplitude and the $\rho(1450)\pi $ and 
$\omega\pi^0$ amplitudes are obtained. The $\rho(1450)\pi$ cross section has 
a peak in the energy region of the $\omega(1650)$ resonance (1.55-1.75 GeV).
In this energy range the contributions of the $\rho(770)\pi$ and 
$\rho(1450)\pi$ states are of the same order of magnitude. No resonance 
structure near 1.65 GeV is observed in the $\rho(770)\pi$ cross section.  We
conclude that the intermediate state $\rho(1450)\pi$ gives a significant 
contribution to the decay of $\omega (1650)\to\pi^+\pi^-\pi^0$, whereas the
$\rho(770)\pi$ mechanism dominates in the decay
$\omega(1420)\to\pi^+\pi^-\pi^0$.}

\maketitle

\section{Introduction}
The process $e^+e^-\to\pi^+\pi^-\pi^0$ was studied in many experiments. 
It was first observed in 1969 at the ACO $e^+e^-$ collider~\cite{omega}
when scanning the energy region near the $\omega(782)$ resonance.
Currently the $e^+ e^-\to\pi^+\pi^-\pi^0$ cross section
is measured in detail in the center-of-mass (c.m.) energy ($\sqrt{s}$) range
from 0.6 GeV to 3 GeV. The most accurate data were obtained in the 
SND~\cite{snd1,snd2,snd3,snd4}, CMD-2~\cite{cmd1,cmd2}, and 
BABAR ~\cite{BABAR} experiments. At higher energies, there are the measurements
of the $J/\psi\to\pi^+\pi^-\pi^0$ and $\psi(2S)\to\pi^+\pi^-\pi^0 $ 
decays~\cite{pdg}, and the cross section at $\sqrt {s}=3.67$ and 
$3.77$ GeV in the CLEO~\cite{CLEO} experiment.

{\tolerance=1000
It is usually assumed that the transition through the $\rho(770)\pi$ 
intermediate state dominates in the process $e^+e^-\to\pi^+\pi^-\pi^0$.
Quantitative verification of this assumption was made only in
resonances. In Ref.~\cite{dalitz-om}, the Dalitz plot distribution for
the $\omega\to 3\pi$ decay was analyzed. It was shown that the
distribution is consistent with that for the $\rho(770)\pi$ mechanism.
In Ref.~\cite{dalitz-phi}, the fraction of the $\phi\to 3\pi$ decays
proceeding through the $\rho(770)\pi$ intermediate state was
determined to be $f_{\rho\pi}=94\%$. The fraction of the so-called ``direct 
mechanism'', which can be interpreted also as the transition through the 
$\rho(1450)\pi$ intermediate state, was found to be about 1\%. The rest is 
the interference between these two amplitudes.

}
In the decay of $J /\psi\to\pi^+\pi^-\pi^0$~\cite{dalitz-jpsi}, 
the contribution of the $\rho(1450)\pi$ mechanism increases up to 11\%, 
and $f_{\rho\pi}\approx 114\%$. The interference between the two amplitudes 
is destructive in this decay. The decay $\psi (2S)\to\pi^+\pi^-\pi^0 $ 
has an unusually low branching fraction, 
$(2.01\pm0.17)\times 10^{- 4}$~\cite{pdg}, which is an order of magnitude
less than the estimate made from the $J/\psi$ decay: 
$B(\psi(2S)\to\pi^+\pi^-\pi^0)\approx B(J /\psi\to\pi^+\pi^-\pi^0)
B (\psi(2S)\to e^+ e^-)/B(J/\psi\to e^+ e^-) = 2.8\times 10^{- 3}$.
Also unusual is the Dalitz plot distribution for this
decay~\cite{dalitz-psi2s}. Most events are located in the center of
the distribution, and the two-pion mass spectrum has a wide maximum near 
2.2 GeV. The fraction of events containing $\rho(770)$ is a few percent.

Therefore, it seems interesting to study the dynamics of the process
$e^+ e^-\to\pi^+\pi^-\pi^0$ in the region $\sqrt{s} =1.1$--2.0 GeV, where
there are two excited resonances of the $\omega$ family: $\omega(1420)$
and $\omega(1650)$.
Two-pion invariant mass spectra for this energy region
are given in Refs.~\cite{snd2,BABAR}. In the $\pi^+\pi^-$ mass spectrum
a narrow peak is seen near the $\omega(782)$ mass, which is
explained by the contribution of the process $e^+e^-\to\omega\pi^0$ with
the $\omega$ decaying to $\pi^+\pi^-$. This phenomenon 
was predicted theoretically in Ref.~\cite{tomp3pi-1}. 
In the energy range 1.1--1.4 GeV, two-pion mass spectra are well 
described by the sum of the $\rho(770)\pi$ and $\omega\pi^0$ intermediate
states~\cite{snd2,BABAR}. However, in the range $\sqrt{s}=1.4$--2.0 GeV 
a significant deviation from this model is observed in the $\pi^\pm\pi^0$
mass spectrum, which reveals in a shift of the $\rho$-meson peak position
and a bump at mass about 1 GeV. In Ref.~\cite{BABAR}, the contribution
of the $\omega(1650)\to\rho(1450)\pi$ decay, which interferes
with the  $\rho(770)\pi$ amplitude, is suggested as a 
possible explanation for this deviation.

The main goal of this work is to study the dynamics of the process
$e^+e^-\to\pi^+\pi^-\pi^0 $ in the energy range from 1.15 GeV to 2.00 GeV
using data accumulated in the SND experiment at the VEPP-2000 $e^+e^-$
collider~\cite{vepp2k}.

\section{Detector and experiment}
{\tolerance=300
The Spherical Neutral Detector (SND) is an universal nonmagnetic
detector collecting data at the VEPP-2000 $e^+e^-$ collider.
A detailed description of detector subsystems can be found in
Refs.~\cite{SNDdet}. The main part of the detector is the three-layer 
spherical electromagnetic calorimeter based on  NaI (Tl) crystals. 
The calorimeter covers 95\% of the solid angle. Its energy resolution for
photons is $\sigma_E/E = 4.2\%/\sqrt[4]{E(\mbox{GeV})}$, and the angular
resolution (r.m.s) is about 1.5$^{\circ}$. Parameters of charged particles are
measured using a nine-layer drift chamber and a single-layer
proportional chamber with cathode strip readout located in a common gas volume.
The solid angle of the tracking system is 94\% of $4\pi$. Its angular
resolution (r.m.s) is $0.45^\circ$ and $0.8^\circ$ for the azimuthal and polar
angles, respectively. The muon system is located outside the calorimeter and 
consists of proportional tubes and scintillation counters.

}
The analysis is based on data recorded in the SND experiment in 2011 and 
2012. Several scans of the energy region from 1.05 to 2.00 GeV with a total
integrated luminosity of 34~pb$^{-1}$ were performed with a step of 20--25 MeV.
The 2011 data set was used previously~\cite{snd4} to measure 
the $e^+e^-\to\pi^+\pi^-\pi^0$ cross section.

The luminosity in this analysis is measured using the process of elastic
scattering $e^+e^-\to e^+e^-$ with an accuracy better than 2\%~\cite{snd4}.

\section{\boldmath Event selection and measurement of the
$e^+e^-\to\pi^+\pi^-\pi^0$ cross section\label{sel}}
\begin{figure}
\centering
\resizebox{0.45\textwidth}{!}{\includegraphics{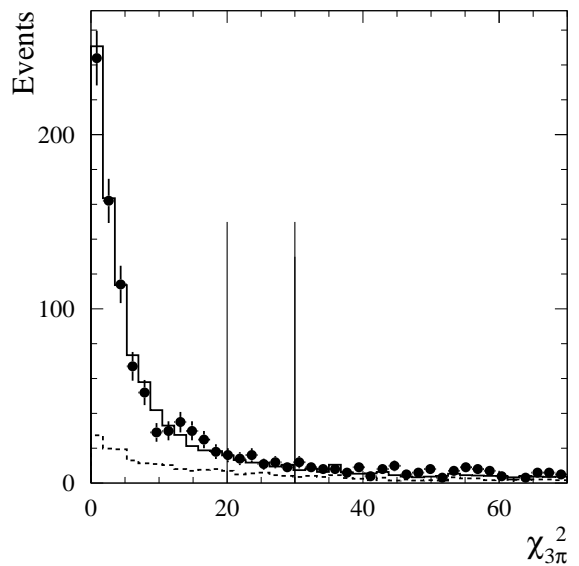}}
\caption{ The $\chi^2_{3\pi}$ distribution for data events
with $\sqrt{s}=1.44$ GeV (points with error bars). The
solid curve is a sum of simulated distributions for signal and background
events. The dashed curve shows the background contribution. The simulated 
distributions are normalized to the numbers of signal and background events
determined from the fit to the two-photon invariant mass distribution.
The lines indicate the cuts used for cross section measurement 
($\chi^2_{3\pi}<30$) and for dynamics study ($\chi^2_{3\pi}<20$).
\label{hi2dis}}
\end{figure}

{\tolerance=300
The selection of $e^+ e^-\to\pi^+\pi^-\pi^0 $ candidate events is described
in detail in Ref.~\cite {snd4}. The following criteria are applied. 
The candidate event contains two charged particles originating from the
beam interaction region and two photons with energy higher than 30 MeV. 
The polar angles of the charged particles must be in the range from 
$30^\circ$ to $150^\circ $. Background from the two-body processes 
$e^+e^-\to e^+e^-$, $\mu^+\mu^-$, $\pi^+\pi^-$, and $K^+K^-$ is rejected by
the condition $|180^\circ-|\varphi_1-\varphi_2||>10^\circ$, where
$\varphi_i$ are the charged-particle azimuthal angles.
To suppress beam-generated background and background from QED processes,
e.g., $e^+e^-\to e^+e^-\gamma$, the condition on the total energy deposition 
in the calorimeter $0.3<E_{\rm tot}/\sqrt{s}<0.8 $ is applied. 
The QED processes are additionally suppressed by the requirement that the 
energy deposition in the calorimeter from charged particles is less 
than $0.6\sqrt{s}$. 
}
For events passing the selection criteria described above, the vertex fit 
is performed using parameters of two charged tracks. 
The found vertex is used to refine the parameters of charged particles
and photons. Then the kinematic fit
to the hypothesis $e^+ e^-\to\pi^+\pi^-\gamma\gamma $ is performed with
the four constraints of energy and momentum conservation.
As a result of the fit, the momenta of charged particles are determined, and
the photon energies and angles are refined. The quality of the fit is
characterized by the parameter $\chi^2_{3\pi}$. The $\chi^2_{3\pi}$ 
distribution for data events is compared with the simulated signal+background
distribution in Fig.~\ref{hi2dis}. 
\begin{figure}
\centering
\resizebox{0.45\textwidth}{!}{\includegraphics{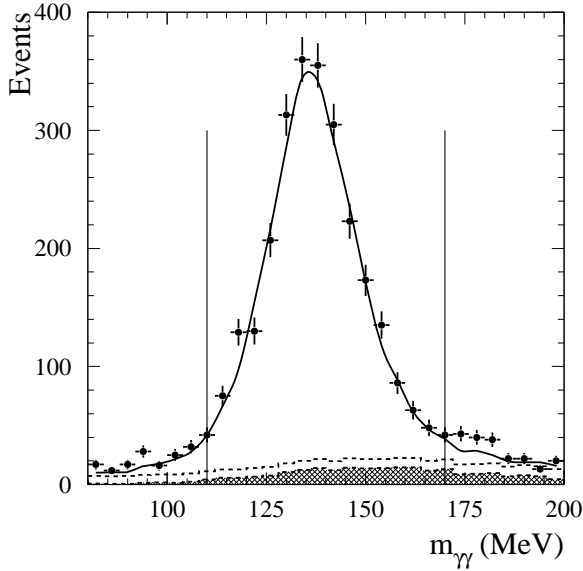}}
\caption{The two-photon invariant mass distribution for selected data events
from the energy region $\sqrt{s}=1.28$--1.52 GeV (points with error bars). The
solid curve is the result of the fit described in the text. The dashed curve 
shows the total background contribution. The hatched histogram is the
distribution for $e^+e^- \to \pi^+\pi^-\pi^0\pi^0$ background events.
The lines indicate the boundaries of the mass window used in the dynamics 
study. 
\label{mgg-spec}}
\end{figure}
Finally, we select events with $\chi^2_{3\pi}<30$ and analyze 
the two-photon invariant mass ($m_{\gamma\gamma}$) distribution.
This distribution for four energy points of the 2012 scan 
($\sqrt{s} = 1.28$-1.52 GeV) is shown in Fig.~\ref{mgg-spec}. It is fitted
with a sum of signal and background distributions. The signal distribution is
obtained using $e^+e^- \to \pi^+\pi^-\pi^0$ simulation.
\begin{table*}
\centering
\caption{The c.m. energy ($\sqrt{s}$), integrated luminosity ($L$),
number of signal events ($N_{3\pi}$), detection efficiency
($\varepsilon$), radiative correction factor ($1+\delta$), and Born cross
section ($\sigma$) for 15 energy points of the 2012 scan.
For $N_{3\pi}$ the statistical error is quoted. For the cross section the 
first error is statistical, the second is systematic.
\label{result2012}}
\renewcommand{\tabcolsep}{0.5cm}
\begin{tabular}{cccccc}
\hline
{$\sqrt{s}$, GeV } & {$L$, nb$^{-1}$}  & {$N_{3\pi}$} & 
{$\varepsilon$,\%} & {$1+\delta$} & {$\sigma$, nb} \\ \hline 
1.28 & 759.5 & $679.2\pm 35.0$& 18.77 & .9123 &  5.22  $\pm$  0.27  $\pm$  0.23\\
1.36 & 837.4 & $638.2\pm 34.7$& 18.77 & .9235 &  4.40  $\pm$  0.24  $\pm$  0.19\\
1.44 & 1015.6& $713.4\pm 35.9$& 19.07 & .9132 &  4.03  $\pm$  0.20  $\pm$  0.18\\
1.52 & 670.3 & $498.6\pm 32.6$& 19.07 & .8977 &  4.34  $\pm$  0.28  $\pm$  0.19\\
1.68 & 903.2 & $580.5\pm 34.8$& 19.00 & .9409 &  3.60  $\pm$  0.22  $\pm$  0.16\\
1.72 & 503.6 & $211.1\pm 26.1$& 18.18 & .9733 &  2.37  $\pm$  0.29  $\pm$  0.10\\
1.76 & 894.3 & $291.6\pm 29.5$& 18.18 & .9906 &  1.81  $\pm$  0.18  $\pm$  0.08\\
1.80  & 982.3 & $206.9\pm 26.2$& 17.97 & .9974 &  1.18  $\pm$  0.15  $\pm$  0.05\\
1.84 & 781.9 & $146.1\pm 14.9$& 17.70 & .9874 &  1.07  $\pm$  0.11  $\pm$  0.05\\
1.872& 919.4 & $153.0\pm 21.7$& 16.85 & .9815 &  1.01  $\pm$  0.14  $\pm$  0.04\\
1.90  & 943.3 & $63.4 \pm 26.5$& 16.45 & .9628 &  0.42  $\pm$  0.18  $\pm$  0.02\\
1.92 & 659.5 & $60.3 \pm 20.0$& 16.53 & .9429 &  0.59  $\pm$  0.19  $\pm$  0.03\\
1.94 & 923.9 & $132.5\pm 24.4$& 15.98 & .9401 &  0.95  $\pm$  0.18  $\pm$  0.04\\
1.96 & 724.0 & $52.0 \pm 17.4$& 15.85 & .9369 &  0.48  $\pm$  0.16  $\pm$  0.02\\
1.98 & 637.1 & $56.7 \pm 15.3$& 15.42 & .9252 &  0.62  $\pm$  0.17  $\pm$  0.03\\
\hline
\end{tabular}
\end{table*}

The main sources of background in the energy region under study are the 
processes $e^+e^-\to\pi^+\pi^-\pi^0\pi^0 $ and $e^+e^-\to\pi^+\pi^-\gamma$.
The first background process has the $m_{\gamma\gamma}$ spectrum with a wide
maximum to the right of the $\pi^0$ peak. The shape of the 
$m_{\gamma\gamma}$ spectrum for $e^+ e^-\to\pi^+\pi^-\gamma$ events as well 
as for other background processes ($e^+e^-\to e^+e^-\gamma $, 
$e^+e^-\to\mu^+\mu^-\gamma$, $e^+e^-\to K^+K^-\pi^0$,{\ldots}) is close to
linear. In the fit, the background is described by the sum
of the simulated distribution for the process $e^+ e^-\to\pi^+\pi^-\pi^0\pi^0$
and a linear function. The fit parameters are the number of signal
events ($N_{3\pi}$), the number of background $e^+ e^-\to\pi^+\pi^-\pi^0\pi^0$
events ($N_{4\pi}$), and parameters of the linear function. The fitted curve
as well as the contributions of the two components of the background are shown 
in Fig.~\ref{mgg-spec}. The fitted $N_{4\pi}$ value is consistent with
the number of $e^+e^-\to\pi^+\pi^-\pi^0\pi^0$ events expected from
simulation.
\begin{figure}
\centering
\resizebox{0.45\textwidth}{!}{\includegraphics{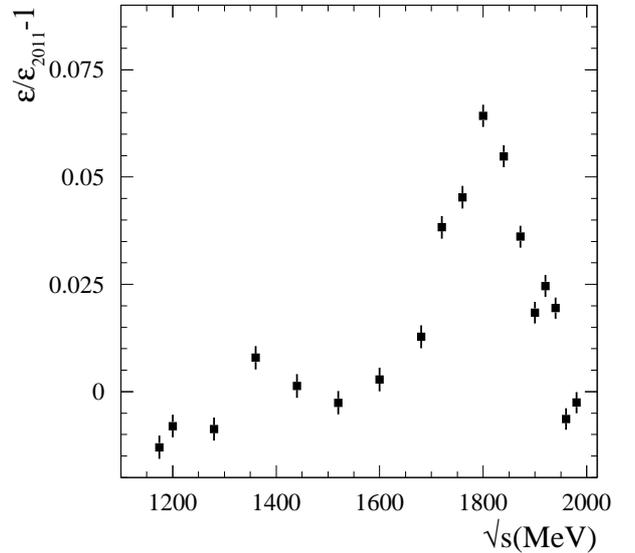}}
\caption{The relative difference between the detection efficiency for 
$e^+e^-\to\pi^+\pi^-\pi^0$ events calculated in this work 
($\varepsilon$) and the efficiency calculated in Ref.~\cite{snd4} 
($\varepsilon_{2011}$) using the model from Ref.~\cite{Czyz}.
\label{effcor}}
\end{figure}

The fitted numbers of signal events for the 2012 scan are listed in 
Table~\ref{result2012} together with the integrated luminosity $L$,
detection efficiency $\varepsilon$, and radiative correction $1+\delta$.
The detection efficiency is calculated using the Monte-Carlo simulation
performed in the model defined below in Sec.~\ref{dynanal}. The model
includes the intermediate states $\rho(770)\pi$, $\rho(1450)\pi$, and
$\omega\pi^0$. Its parameters are determined in Sec.~\ref{dynanal} from a fit
to the Dalitz plot distribution and the $\pi^+\pi^-$ mass spectrum
for data events. The model uncertainty of the detection 
efficiency is estimated by variation of the model parameters within their
errors and does not exceed 1\%. The detection efficiency is also corrected 
for the difference between data and simulation in the $\chi^2_{3\pi}$ 
distribution and the number of photons in an event. This difference was
studied in Ref.~\cite{snd4} and was found to be $(1.9\pm3.1)\%$.

The radiative correction factor is calculated during the fit to the visible 
cross section data ($N_{3\pi}/L$) with the vector-meson-dominance (VMD) model,
as described in Ref.~\cite{snd4}. The Born cross section is then calculated
as $\sigma=N_{3\pi}/[\varepsilon L (1+\delta)]$. 

The detailed analysis of systematic uncertainties on the measured cross 
section was carried out in Ref.~\cite{snd4}. The total systematic uncertainty
is 4.4\% and includes the uncertainties in the luminosity measurement(2\%), 
the detection efficiency (3.1\%), the numbers of signal events (2\%), 
the radiative correction (1\%), and the model error mentioned above (1\%).

In the analysis of the 2011 data set~\cite{snd4}, the detection efficiency was
determined using the simulation based on the model from
Ref.~\cite{Czyz}. This model includes the $\rho(770)\pi$, $\rho(1450)\pi$,
$\rho(1700)\pi$, and $\omega\pi^0$ intermediate states. Its parameters
are chosen to reproduce the measured energy dependence of the 
$e^+ e^-\to\pi^+\pi^-\pi^0$ cross section and the two-pion invariant mass 
spectra from Ref.~\cite{BABAR}. The relative difference between
the detection efficiencies obtained in our model ($\varepsilon$) and 
in the model~\cite{Czyz} ($\varepsilon_{2011}$) as a function of energy is
shown in Fig.~\ref{effcor}. To understand the source of the about 6\%
difference between the models observed near 1.8 GeV, we compare them 
with the pure $\rho(770)\pi$ mechanism. The relative difference in the 
detection efficiency between our model and the $\rho(770)\pi$ 
($\varepsilon/\varepsilon_{\rho\pi}-1$) is in the range between $-2\%$ and
$2.5\%$. Near 1.8 GeV the difference is 1.8\%. For the model from
Ref.~\cite{Czyz} the relative difference 
$\varepsilon_{2011}/\varepsilon_{\rho\pi}-1$ has a a minimum in the
range 1.7--1.9 GeV. Its minimal value is $-4.5\%$ at 1.85 GeV. We conclude that
the main source of the 6\% difference between the models in Fig.~\ref{effcor}
is imperfect description of the intermediate states in the $3\pi$ system
in the model of Ref.~\cite{Czyz}. Taking this difference as an efficiency 
correction, we reanalyze the 2011 data. The corrected cross section values are
listed in Table~\ref{result2011}.
\begin{table*}
\centering
\caption{The c.m. energy ($\sqrt{s}$) and Born cross section 
($\sigma$) for 40 energy points of the 2011 scan. The quoted errors
are statistical and systematic, respectively.
\label{result2011} }
\begin{tabular}{cccccccc}
\hline
{$\sqrt{s}$, GeV}&{$\sigma$, nb}&{$\sqrt{s}$, GeV} & {$\sigma$, nb}&{$\sqrt{s}$, GeV} & {$\sigma$, nb}&{$\sqrt{s}$, GeV} & {$\sigma$, nb} \\
\hline 
1.050 & 1.27 $\pm$ 0.48 $\pm$0.26 & 1.300 & 4.92 $\pm$ 0.26 $\pm$0.22 &1.550 & 4.63 $\pm$ 0.24 $\pm$0.20 & 1.800 & 1.05 $\pm$ 0.18 $\pm$0.05\\ 
1.075 & 3.30 $\pm$ 0.26 $\pm$0.40 & 1.325 & 4.91 $\pm$ 0.22 $\pm$0.22 &1.575 & 4.71 $\pm$ 0.24 $\pm$0.21 & 1.825 & 1.28 $\pm$ 0.14 $\pm$0.06\\
1.100 & 4.27 $\pm$ 0.32 $\pm$0.34 & 1.350 & 5.02 $\pm$ 0.24 $\pm$0.22 &1.600 & 5.81 $\pm$ 0.27 $\pm$0.26 & 1.850 & 1.28 $\pm$ 0.17 $\pm$0.06\\ 
1.125 & 4.64 $\pm$ 0.26 $\pm$0.32 & 1.375 & 4.81 $\pm$ 0.22 $\pm$0.21 &1.625 & 5.06 $\pm$ 0.28 $\pm$0.22 & 1.870 & 0.92 $\pm$ 0.13 $\pm$0.04\\ 
1.150 & 5.24 $\pm$ 0.29 $\pm$0.31 & 1.400 & 4.18 $\pm$ 0.24 $\pm$0.18 &1.650 & 4.65 $\pm$ 0.26 $\pm$0.20 & 1.890 & 0.68 $\pm$ 0.12 $\pm$0.03\\ 
1.175 & 5.42 $\pm$ 0.27 $\pm$0.24 & 1.425 & 4.06 $\pm$ 0.23 $\pm$0.18 &1.675 & 3.42 $\pm$ 0.22 $\pm$0.15 & 1.900 & 1.04 $\pm$ 0.15 $\pm$0.05\\ 
1.200 & 5.13 $\pm$ 0.28 $\pm$0.23 & 1.450 & 4.10 $\pm$ 0.25 $\pm$0.18 &1.700 & 2.61 $\pm$ 0.23 $\pm$0.12 & 1.925 & 0.66 $\pm$ 0.11 $\pm$0.03\\ 
1.225 & 5.80 $\pm$ 0.27 $\pm$0.26 & 1.475 & 4.30 $\pm$ 0.21 $\pm$0.19 &1.725 & 2.15 $\pm$ 0.19 $\pm$0.09 & 1.950 & 0.51 $\pm$ 0.13 $\pm$0.02\\ 
1.250 & 6.00 $\pm$ 0.28 $\pm$0.26 & 1.500 & 4.44 $\pm$ 0.19 $\pm$0.20 &1.750 & 1.80 $\pm$ 0.18 $\pm$0.08 & 1.975 & 0.69 $\pm$ 0.14 $\pm$0.03\\ 
1.275 & 5.55 $\pm$ 0.29 $\pm$0.24 & 1.525 & 4.52 $\pm$ 0.24 $\pm$0.20 &1.775 & 1.62 $\pm$ 0.16 $\pm$0.07 & 2.000 & 0.84 $\pm$ 0.16 $\pm$0.04\\ 
\hline
\end{tabular}
\end{table*}
\begin{figure*}
\centering
\resizebox{0.7\textwidth}{!}{\includegraphics{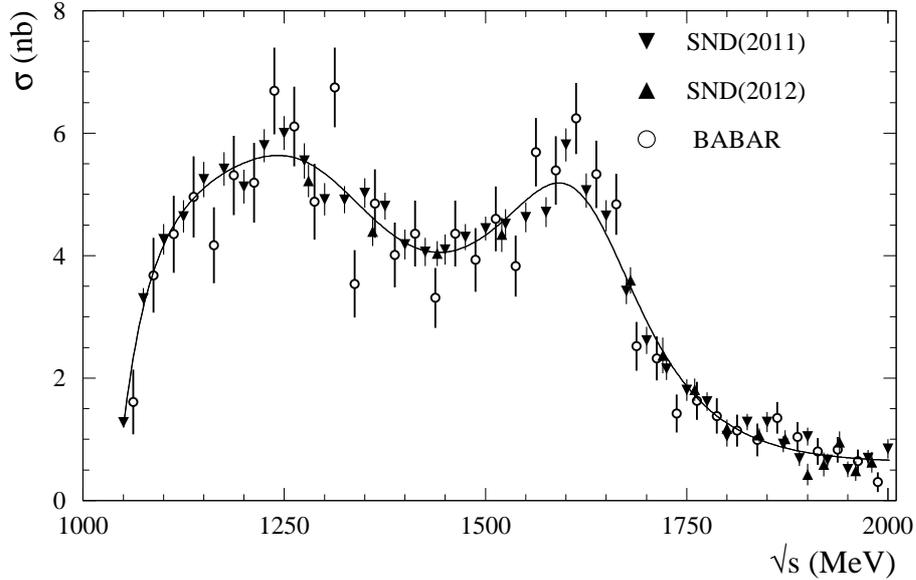}}
\caption{The Born cross section for the process $e^+e^- \to \pi^+\pi^-\pi^0$
measured in this work for 2011 and 2012 scans, in comparison
with the results of the BABAR experiment~\cite{BABAR}. 
The curve represents the result of the fit to the SND data with a sum of 
contributions from the resonances $\omega(782)$, $\phi(1020)$, $\omega(1420)$, 
and $\omega(1650)$.
\label{crs-res}}
\end{figure*}

The $e^+e^- \to \pi^+\pi^-\pi^0$ cross section obtained in this work
in comparison with the BABAR measurement ~\cite{BABAR}, as well as the result
of the fit to the SND data with a sum of contributions of isoscalar 
resonances~\cite{snd4} are shown in Fig.~\ref{crs-res}. It is seen 
that the two SND measurements are in good agreement
with each other and with the result of BABAR~\cite{BABAR}. The two peaks
in the cross section correspond to the $\omega(1420)$ and $\omega(1650)$
resonances.

\section{\boldmath Dynamics of the process $e^+e^- \to \pi^+\pi^-\pi^0$}
\label{dynanal}
\begin{table*}
\centering
\caption{The c.m. energy interval ($\sqrt{s}$), number of selected 
$e^+e^-\to \pi^+\pi^-\pi^0$ events ($N_{3\pi}$), number of background events
($N_{\rm bkg}$),  cross sections for intermediate states 
$\rho(770)\pi$ ($\sigma_{\rho\pi}$),
$\rho(1450)\pi$ ($\sigma_{\rho^\prime\pi}$), and $\omega\pi^0$ 
($\sigma_{\omega\pi}$), and relative phases between the amplitudes of the 
intermediate states $\rho(1450)\pi$ and $\rho(770)\pi$ ($\phi_1$), and  
$\omega\pi^0$ and $\rho(770)\pi$ ($\phi_2$).\label{tab2}}
\renewcommand{\arraystretch}{1.3}
\renewcommand{\tabcolsep}{0.5cm}
\begin{tabular}{cccccccc}
\hline
{$\sqrt{s}$, GeV } & {$N_{3\pi}$} & $N_{\rm bkg}$ & $\sigma_{\rho\pi}$, nb &
$\sigma_{\rho^\prime\pi}$, nb & $\sigma_{\omega\pi}$, nb & $\phi_1$, rad &
$\phi_2$, rad \\ \hline
1.15--1.18 & $ 957\pm  31$ & 266 & $4.40_{-0.26}^{+0.48}$ &$0.05_{-0.07}^{+0.07}$ & $0.21\pm 0.03$ & -- &$2.02_{-0.48}^{+0.50}$\\
1.20--1.23 & $1067\pm  33$ & 128 & $4.68_{-0.24}^{+0.32}$ &$0.01_{-0.03}^{+0.03}$ & $0.17\pm 0.02$ & -- &$1.54_{-0.39}^{+0.36}$\\
1.25--1.30 & $2021\pm  45$ & 241 & $4.25_{-0.15}^{+0.22}$ &$0.06_{-0.09}^{+0.09}$ & $0.22\pm 0.02$ & -- &$1.28_{-0.23}^{+0.22}$\\
1.32--1.38 & $1642\pm  41$ & 201 & $4.29_{-0.22}^{+0.18}$ &$0.06_{-0.04}^{+0.05}$ & $0.26\pm 0.03$ & -- &$2.26_{-0.23}^{+0.21}$\\
1.42--1.48 & $1631\pm  40$ & 217 & $3.43_{-0.28}^{+0.25}$ &$0.01_{-0.02}^{+0.01}$ & $0.25\pm 0.03$ & -- &$1.66_{-0.51}^{+0.33}$\\
1.50--1.55 & $1836\pm  43$ & 217 & $2.73_{-0.23}^{+0.23}$ &$0.25_{-0.10}^{+0.10}$ & $0.21\pm 0.03$ & $1.26_{-0.22}^{+0.18}$&$1.82_{-0.28}^{+0.24}$\\
1.57--1.60 & $1679\pm  41$ & 143 & $2.76_{-0.29}^{+0.28}$ &$0.81_{-0.25}^{+0.29}$ & $0.14\pm 0.02$ & $1.80_{-0.18}^{+0.17}$&$2.10_{-0.40}^{+0.31}$\\
1.65--1.68 & $1252\pm  35$ & 115 & $2.12_{-0.23}^{+0.22}$ &$0.87_{-0.23}^{+0.26}$ & $0.08\pm 0.01$ & $2.30_{-0.16}^{+0.17}$&$2.36_{-1.07}^{+0.51}$\\
1.70--1.72 & $ 445\pm  21$ &  48 & $2.02_{-0.26}^{+0.26}$ &$0.48_{-0.18}^{+0.20}$ & $0.06\pm 0.01$ & $2.67_{-0.25}^{+0.18}$&$1.13_{-0.82}^{+1.11}$\\
1.75--1.78 & $ 599\pm  24$ &  84 & $2.00_{-0.25}^{+0.24}$ &$0.27_{-0.09}^{+0.08}$ & $0.03\pm 0.01$ & $3.27_{-0.40}^{+0.35}$&$3.97_{-0.73}^{+0.80}$\\
1.80--1.85 & $ 540\pm  23$ & 105 & $1.20_{-0.24}^{+0.20}$ &$0.19_{-0.07}^{+0.10}$ & $0.03\pm 0.01$ & $3.21_{-0.47}^{+0.45}$&$3.21_{-1.51}^{+1.33}$\\
1.87--1.90 & $ 433\pm  21$ &  95 & $1.14_{-0.15}^{+0.11}$ &$0.18_{-0.08}^{+0.06}$ & $0.02\pm 0.01$ & $3.84_{-0.36}^{+0.17}$&$2.77_{-1.86}^{+0.95}$\\
1.92--1.94 & $ 278\pm  17$ &  64 & $0.30_{-0.11}^{+0.12}$ &$0.17_{-0.13}^{+0.19}$ & $0.01\pm 0.01$ & $1.63_{-0.40}^{+0.44}$&$1.13_{-1.52}^{+1.75}$\\
1.96--2.00 & $ 239\pm  15$ &  58 & $0.32_{-0.10}^{+0.10}$ &$0.09_{-0.30}^{+0.12}$ & $0.01\pm 0.01$ & -- &$0.65_{-1.20}^{+3.53}$\\
\hline
\end{tabular}
\end{table*}
To study the dynamics of the process $e^+e^-\to\pi^+\pi^-\pi^0$, we analyze
the Dalitz plot distribution and the spectrum of the $\pi^+\pi^-$invariant 
mass. Data from 2011 and 2012 scans from the energy range
$\sqrt{s}=1.15$--2.00 GeV are used. The range $\sqrt{s}=1.05$--1.15 GeV,
in which selected events contain significant fraction of radiative-return
$e^+e^-\to\phi(1020)\gamma\to\pi^+\pi^-\pi^0\gamma$ events, is excluded from
the analysis. Data with a total integrated luminosity 
of about 28~pb$^{-1}$ are combined into 14 intervals listed in
Table~\ref{tab2}.

For the Dalitz plot analysis, the event selection criteria
are tightened. In addition to the criteria described in Sec.~\ref{sel}
the condition $\chi^2_{3\pi}<20$ and $110<m_{\gamma\gamma}<170$ MeV are
applied. The numbers of selected signal and background events in 
this $m_{\gamma\gamma}$ range are listed in Table~\ref{tab2} for each energy
interval. They are determined from the fit to the $m_{\gamma\gamma}$ spectrum 
as described in Sec.~\ref{sel}.

To describe the dynamics of the process $e^+e^-\to\pi^+\pi^-\pi^0 $, 
a model is used, in which the differential cross section is presented
as a sum of contributions of the three intermediate states
$\rho(770)\pi$, $\rho(1450)\pi$, and $\omega\pi^0$:
\begin{equation}
\label{f1}
\frac{d\sigma}{d\Gamma} = \left| \alpha A_{\rho\pi}+\beta A_{\rho^\prime\pi}+
\gamma A_{\omega\pi}\right|^2,
\end{equation}
where $d\Gamma$ is a phase space element. 
The amplitudes $A_{\rho\pi}$, $A_{\rho^\prime\pi}$, and $A_{\omega\pi}$
are the functions of $s$ and pions momenta. For example,
$|A_{\rho\pi}|^2$ is proportional to
\begin{equation}
\sin^2{\theta_n} ({\mathbf p_+}\times {\mathbf p_-})^2 
\left| \sum_k \frac{m_{\rho^k}^2}
{q^2_k-m_{\rho^k}^2+i q_k \Gamma_{\rho^k}(q^2_k)} \right|^2,
\label{f2}
\end{equation}
where $\theta_n$ is the angle between the normal to the reaction plane and
the beam axis, ${\mathbf p_+}$ and ${\mathbf p_-}$ are the 
charged-pion momenta,
\begin{equation}
\Gamma_{\rho^k}(q^2_k)=\Gamma_{\rho^k}
\left ( \frac{p_\pi(q^2_k)}{p_\pi(m_{\rho^k}^2)} 
\right )^3
\frac{m^2_k}{q^2_k},
\label{f3}
\end{equation}
$k$ takes values  $+,-,0$, $m_{\rho^k}$ and $\Gamma_{\rho^k}$ are the mass and
width of the $\rho^k(770)$, $q^2_k$ is the invariant mass of the pion pair,
$p_\pi$ is the pion momentum in the $\rho$ rest frame.

The  $\rho(1450)\pi$ amplitude is obtained from the $\rho(770)\pi$ amplitude
by replacing the $\rho(770)$ mass and width with the same parameters for
the $\rho(1450)$. In the $\omega\pi^0$ amplitude, the sum over
the three charge combinations is replaced by
$m_\omega^2/(q^2_0-m_\omega^2+i m_\omega \Gamma_\omega)$, where
$m_\omega$ and $\Gamma_\omega$ are the $\omega$ mass and width.
All resonance parameters used in Eq.~(\ref{f1}) are taken from the
Particle Data Group tables~\cite{pdg}.

Other intermediate states, such as $\rho(1700)\pi$ and the direct transition
$\gamma^\ast\to \pi^+\pi^-\pi^0$, also may contribute to the $e^+e^-\to
\pi^+\pi^-\pi^0$ reaction. The Dalitz plot distributions for these
intermediate states are very similar to that for $\rho(1450)\pi$. At our
level of statistics we cannot separate these three intermediate states.
So, the amplitude $A_{\rho^\prime\pi}$ effectively describes their
total contribution. It should be noted that the $\rho(1700)\pi$ state is 
suppressed by the phase space compared to the $\rho(1450)\pi$. This suppression
is by a factor of about 3 in the the $\omega(1650)$ energy range and grows 
rapidly with decreasing energy.

\begin{figure}
\centering
\resizebox{0.45\textwidth}{!}{\includegraphics{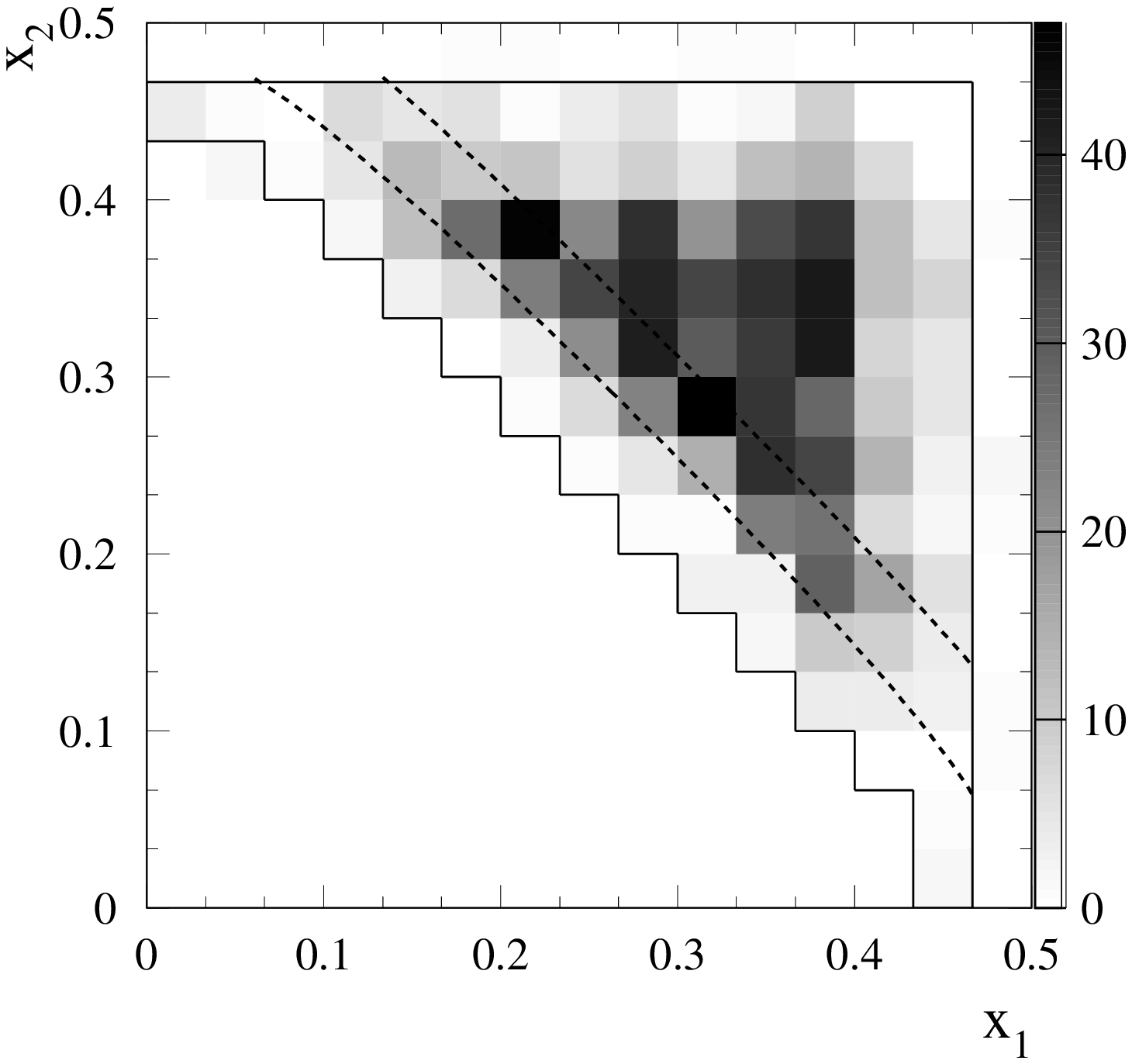}}
\caption{The $x_1$ versus $x_2$ distribution for selected data events from
the interval $\sqrt{s}=1.63-1.68$ GeV. The solid polygon indicate the
Dalitz plot area used in the fit. The area between the dashed curves 
corresponds to the condition $0.68<M_{\pi^+\pi^-}<0.88$ GeV.   
\label{dalitz}}
\end{figure}
The complex coefficients $\alpha$, $\beta$ and $\gamma$ are functions of 
$s$ and are determined from a fit to distributions of kinematic variables.
The Dalitz plot distribution for data events from the interval 
$\sqrt{s}=1.65$--1.68 GeV is shown in Fig.~\ref{dalitz} in
the variables $x_i=p_i/\sqrt{s}$, where $p_i$ ($i=1,2$) are the 
charged pion momenta. Since the signs of charged particles are not
determined in the SND detector, the indices 1 and 2 are assigned randomly. 
We perform a binned fit to the Dalitz plot distribution. The bin size is 
chosen equal $1/30\times 1/30$. The presence of the intermediate mechanism
$\omega\pi^0$ leads to appearance of a narrow structure in the $\pi^+\pi^-$ 
invariant mass ($M_{\pi^+\pi^-}$) spectrum near the $\omega$ mass, 
for description of which the chosen binning is too coarse. Therefore,
events with $0.68<M_{\pi^+\pi^-}<0.88$ GeV are excluded from
the Dalitz plot distribution. A one-dimensional $M_{\pi^+\pi^-}$ distribution
is constructed for them with a 10 MeV bin. The Dalitz plot
distribution and the $M_{\pi^+\pi^-}$ distribution are fitted
simultaneously.

To take into account detector resolution and dependence of the 
detection efficiency on position in the Dalitz plot, the fitting
function for the data Dalitz plot distribution is constructed as follows:
\begin{eqnarray}
\label{difcrs}
D(s,x_1,x_2)& = & \nonumber \\
&&|\alpha|^2 H_{\rho\pi}+|\beta|^2 H_{\rho^\prime\pi}+
|\gamma|^2 H_{\omega\pi}\nonumber\\
&+&2 |\alpha||\beta| \cos(\phi_1) R_{\rho\pi\mbox{-}\rho^\prime\pi}\nonumber \\
&+&2 |\alpha||\beta| \sin(\phi_1) I_{\rho\pi\mbox{-}\rho^\prime\pi}\nonumber \\
&+&2 |\alpha||\gamma| \cos(\phi_2) R_{\rho\pi\mbox{-}\omega\pi}\nonumber \\
&+&2 |\alpha||\gamma| \sin(\phi_2) I_{\rho\pi\mbox{-}\omega\pi}\nonumber \\
&+&2 |\beta||\gamma| \cos(\phi_2-\phi_1) R_{\rho^\prime\pi\mbox{-}\omega\pi}\nonumber \\
&+&2 |\beta||\gamma| \sin(\phi_2-\phi_1) I_{\rho^\prime\pi\mbox{-}\omega\pi},
\end{eqnarray}
where $H_{\rho\pi}$, $H_{\rho^\prime\pi}$, and $H_{\omega\pi}$ are 
the distributions corresponding to the moduli squared of the amplitudes
$A_{\rho\pi}$, $A_{\rho^\prime\pi}$, and $A_{\omega\pi}$, respectively.
They are calculated using MC simulation. For example, to obtain $H_{\rho\pi}$, 
a simulation is performed in the model described by Eq.~(\ref{f1}) 
with $\alpha=1 $ and $\beta=\gamma= 0$.
The simulation takes into account radiation corrections~\cite{radcor},
which are calculated using the Born cross section shown in Fig.~\ref{crs-res}. 
Simulated events pass the selection criteria described above. 
For selected events, a two-dimensional $x_{1}$ versus $x_{2}$ distribution
is constructed. Also the detection efficiency $\varepsilon_{\rho\pi}$
and the cross section
$\sigma_{\rho\pi,{\rm vis}}=(1+\delta)\int |A_{\rho\pi}|^2 d\Gamma$,
where $\delta $ is the radiative correction, are calculated. The
efficiency is corrected for the data-simulation difference as described in
Sec.~\ref{sel}.
The resulting distribution is normalized to the expected number of events
\begin{equation}
\varepsilon_{\rho\pi}(s_i)\sigma_{\rho\pi,{\rm vis}}(s_i)L_i,
\end{equation}
where $L_i$ is the integrated luminosity for the $i$th energy interval.

The distributions $R_{i\mbox{-}j}$ and $I_{i\mbox{-}j}$ correspond to the 
real and imaginary parts of the interference terms 
$A_iA^\ast_j$ ($i\ne j$),
where $i$ and $j$ are $\rho\pi$, $\rho^\prime\pi$, and $\omega\pi$.
To obtain, for example, the distributions $R_{\rho\pi\mbox{-}\rho^\prime\pi}$
and $I_{\rho\pi\mbox{-}\rho^\prime\pi}$, two simulations 
are performed using Eq.~(\ref{f1}) with  $\alpha=1$, $\beta=1$ and
$\gamma=0$, and with $\alpha=1$, $\beta=i$ and $\gamma=0$,
and the distributions $H_{\rho\pi+\rho^\prime\pi}$ and
$H_{\rho\pi+i\rho^\prime\pi}$ are constructed, as it is described 
above for $H_{\rho\pi}$. Then we determine
\begin{eqnarray}
2R_{\rho\pi\mbox{-}\rho^\prime\pi}&=&H_{\rho\pi+\rho^\prime\pi}-
H_{\rho\pi}-H_{\rho^\prime\pi}, \nonumber \\
2I_{\rho\pi\mbox{-}\rho^\prime\pi}&=&H_{\rho\pi+i\rho^\prime\pi}-
H_{\rho\pi}-H_{\rho^\prime\pi}.
\end{eqnarray}
The distributions with the indices $\rho\pi\mbox{-}\omega\pi$ and
$\rho^\prime\pi\mbox{-}\omega\pi$ are built in the same way.
The same technique is used for the $M_{\pi^+\pi^-}$ distribution.

The distributions for background events are obtained using simulation
of the processes $e^+e^- \to \pi^+\pi^-\pi^0\pi^0$~\cite{Czyz-4pi} and $e^+e^-
\to\pi^+\pi^-\gamma$. These two processes produce about 80\% of
background events. The simulated distributions are normalized to 
the number of background events listed in Table~\ref{tab2}. 
It is tested that this background model describes well 
the distribution of two-pion masses for the control regions
$80<m_{\gamma\gamma}<110$ MeV and $170< m_{\gamma\gamma}<200$ MeV.

Due to the interference between the intermediate states $\rho(770)\pi$ and
$\omega\pi^0$, the $M_{\pi^+\pi^-}$ spectrum has a narrow peak-dip structure
near the $\omega$ mass (see, for example \cite{tomp3pi-1}). The shape
of this structure depends on the phase $\phi_2$. The detector resolution
smears the interference pattern. Therefore, only a peak is observed in the
experimental spectrum. This leads to a very strong correlation between the
parameters $|\gamma|$ and $\phi_2$ extracted from the fit to 
the $\pi^+\pi^-$ mass spectrum. The parameter $|\gamma|$ can, however, 
be determined from the Born cross section of the process
$e^+e^-\to\omega\pi^0\to \pi^0\pi^0\gamma$ ($\sigma_{\pi^0\pi^0\gamma}$)
measured by SND~\cite {SNDompi}:
\begin{eqnarray}
\sigma_{\omega\pi}(s_i)&=&|\gamma(s_i)|^2\int |A_{\omega\pi}(s_i)|^2
d\Gamma\nonumber\\
&=&\sigma_{\pi^0\pi^0\gamma}(s_i)\frac{B(\omega\to
\pi^+\pi^-)}{B(\omega\to\pi^0\gamma)},
\label{nompi}
\end{eqnarray}
where $B(\omega\to \pi^+\pi^-)$ and $B(\omega\to\pi^0\gamma)$ are the 
branching fractions of the corresponding $\omega$ decays~\cite{pdg}. 
The values of the cross section $\sigma_{\omega\pi}$ obtained 
using Eq.~(\ref{nompi}) are given in Table~\ref{tab2}.
During the fit the parameter $|\gamma|$ is allowed to vary within
its errors near the calculated value. 

Instead of the parameters $|\alpha|$ and $|\beta|$, we use 
the Born cross sections for the $\rho(770)\pi$ and 
$\rho(1450)\pi$ mechanisms:
\begin{eqnarray}
\sigma_{\rho\pi}(s_i)&=&|\alpha(s_i)|^2\int |A_{\rho\pi}(s_i)|^2
d\Gamma, \nonumber \\
\sigma_{\rho^\prime\pi}(s_i)&=&|\beta(s_i)|^2\int |A_{\rho^\prime\pi}(s_i)|^2
d\Gamma. 
\label{nrhopi}
\end{eqnarray}
These cross sections, as well as the relative phases $\phi_1$ and $\phi_2$, 
are determined from the fit to the Dalitz plot distribution and the
$M_{\pi^+\pi^-}$ spectrum.
\begin{figure*}
\resizebox{0.48\textwidth}{!}{\includegraphics{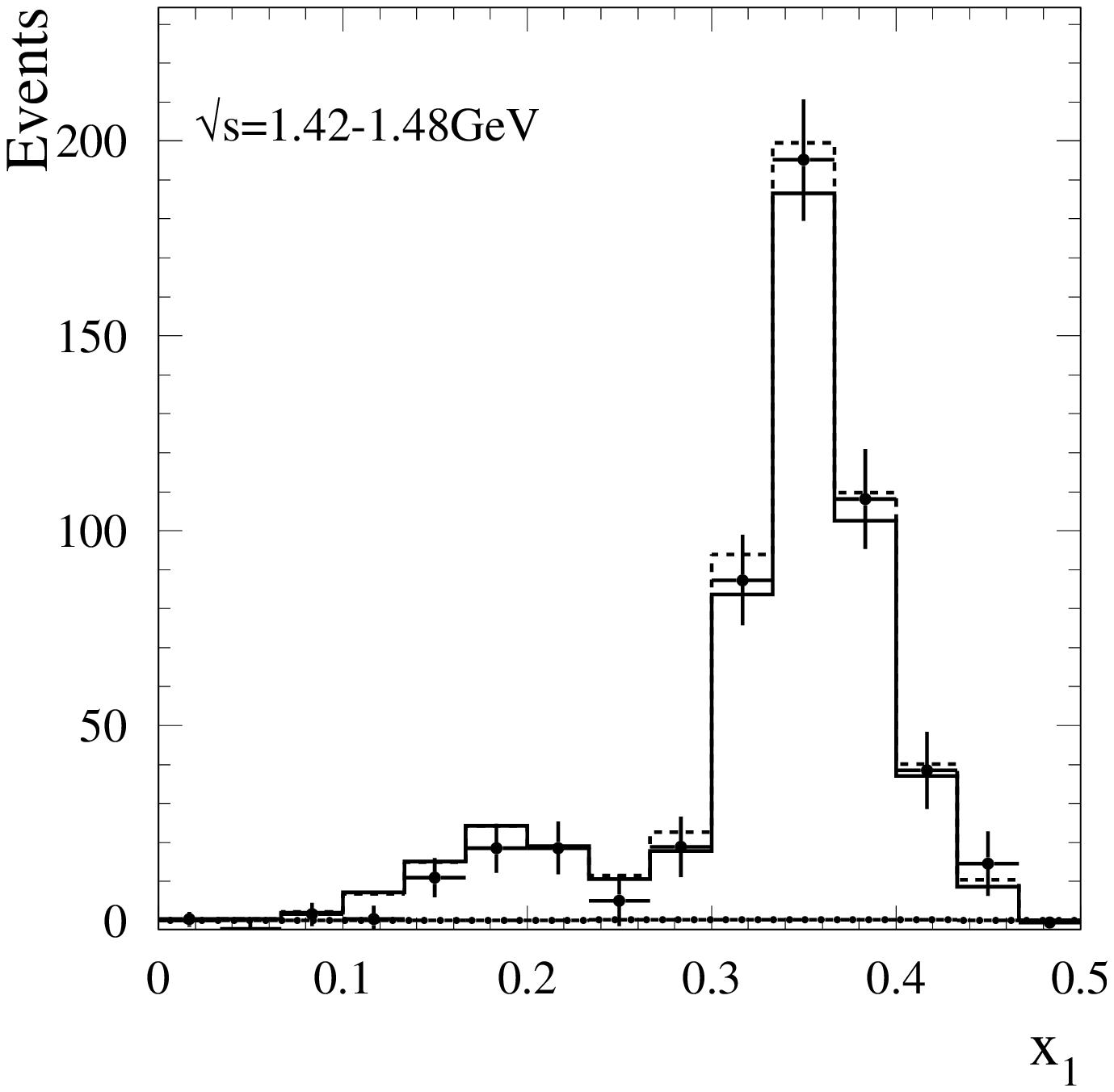}}\hfill
\resizebox{0.48\textwidth}{!}{\includegraphics{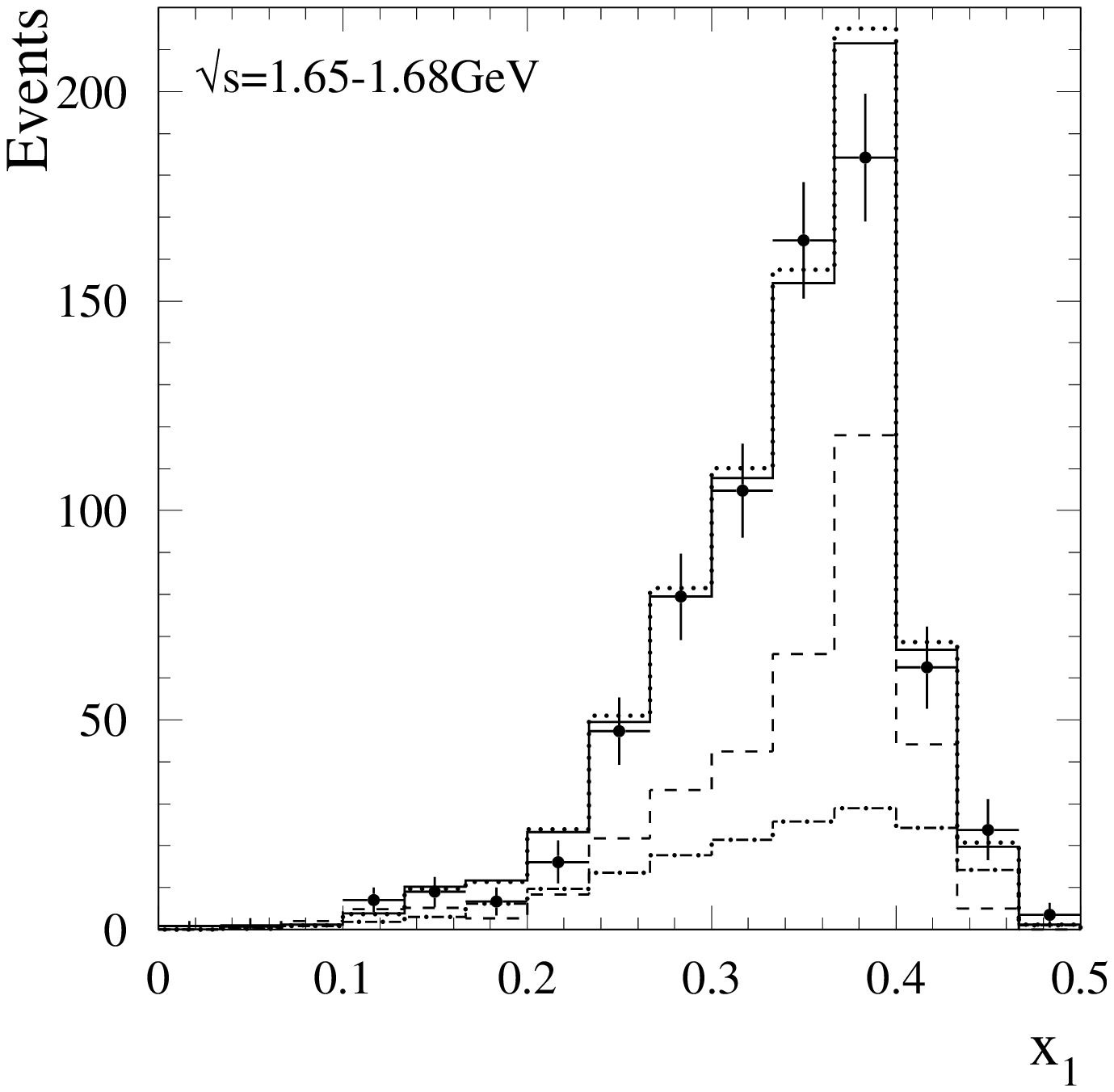}}
\resizebox{0.48\textwidth}{!}{\includegraphics{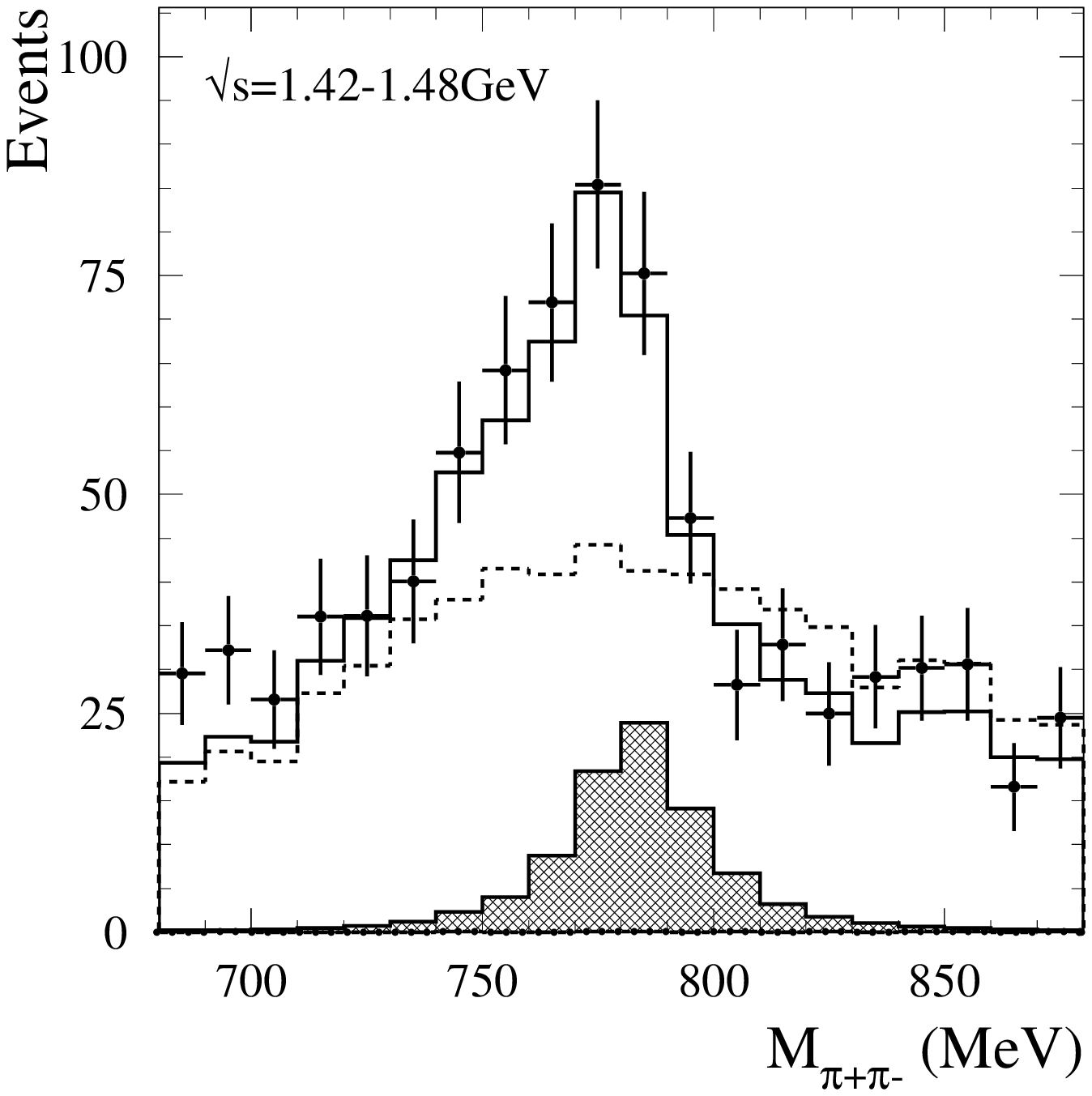}}\hfill
\resizebox{0.48\textwidth}{!}{\includegraphics{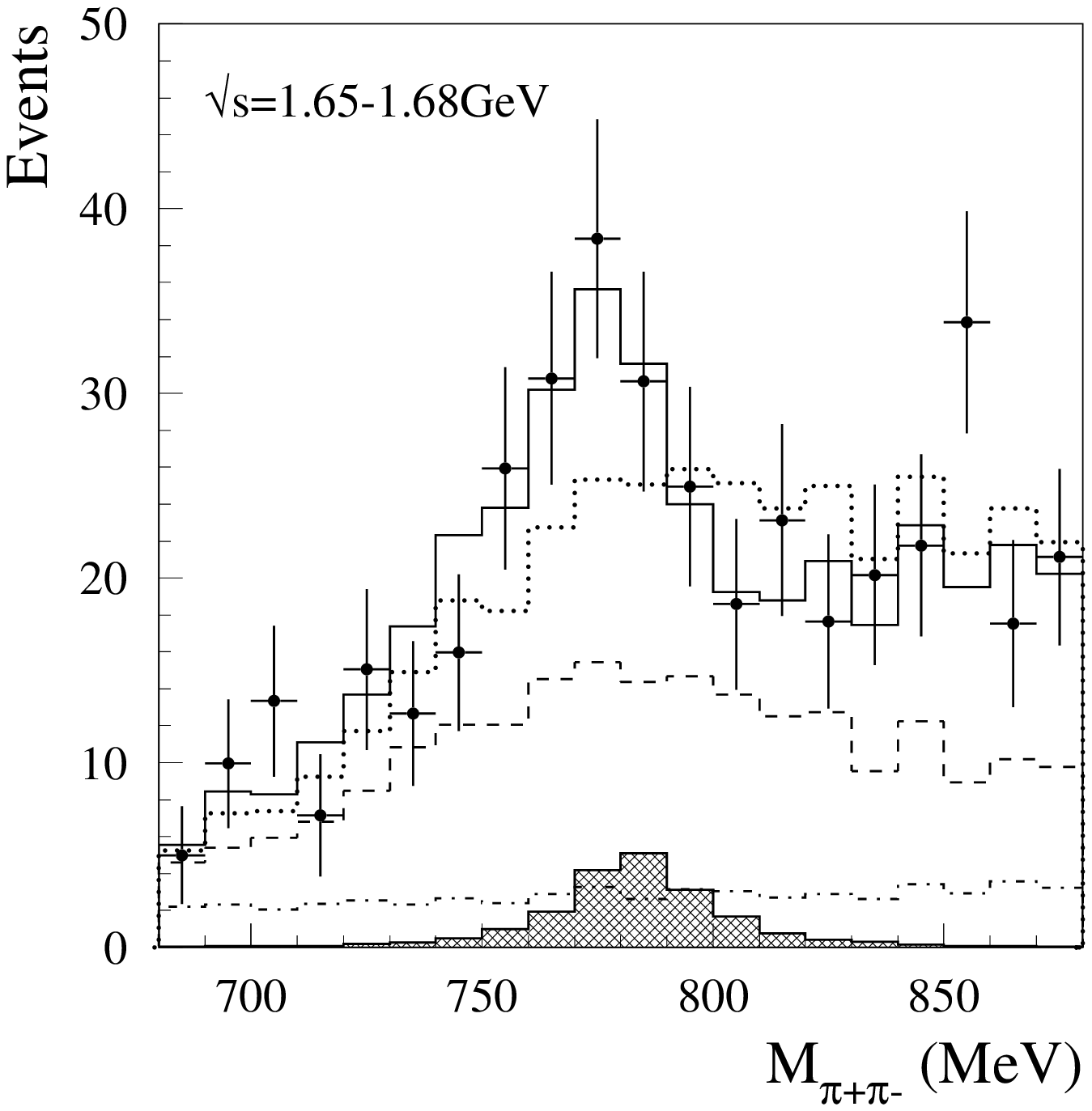}}
\caption{The $x_1$ distribution and $M_{\pi^+\pi^-}$ spectrum 
for two energy intervals: $\sqrt{s}=1.42$--1.48 MeV and
$\sqrt{s}=1.65$--1.68 MeV. The points with error bars represent data.
The solid histogram is the result of the fit described in the text.
The dashed, dash-dotted, and hatched histograms show the distributions
corresponding to the squared amplitudes for the intermediate states
$\rho(770)\pi$, $\rho(1450)\pi$ and $\omega\pi^0$, respectively. The dotted
histograms in the right plots show the total contributions of the
$\rho(770)\pi$ and $\rho(1450)\pi$ intermediate states including 
the interference term.
\label{spectr}}
\end{figure*}

The data $x_1$ and $M_{\pi^+\pi^-}$ distributions 
for the energy intervals $\sqrt{s}=1.42$--1.48 MeV and
$\sqrt{s}=1.65$--1.68 MeV are shown in Fig.~\ref{spectr}.
The distributions obtained as a result of the fit are also shown
together with the spectra corresponding to the squares of the
$\rho(770)\pi$, $\rho(1450)\pi$, and $\omega\pi^0$ amplitudes 
($|\alpha|^2 H_{\rho\pi}$,  $|\beta|^2 H_{\rho^\prime\pi}$,
and $|\gamma|^2 H_{\omega\pi}$).
The relative fraction of the intermediate mechanism, for example $\rho\pi$, 
is defined as follows:
\begin{equation}
f_{\rho\pi}=\frac{\int d\Gamma |\alpha A_{\rho\pi}|^2}
{\int d\Gamma \left| \alpha A_{\rho\pi}+\beta A_{\rho^\prime\pi}+
\gamma A_{\omega\pi}\right|^2}.\label{frac}
\end{equation}
At $\sqrt{s}\approx 1.45$ GeV $f_{\rho\pi}=(84\pm 7)\%$,
$f_{\rho^\prime\pi}=(0.2\pm 0.4)\%$, and $f_{\omega\pi}=(6\pm 1)\%$.
The interference between the isovector ($\omega\pi^0$) and isoscalar
($\rho\pi+\rho^\prime\pi$) amplitudes give a 10\% contribution to the 
total $e^+e^-\to\pi^+\pi^-\pi^0$ cross section. Thus, the
total contribution associated with the intermediate state $\omega\pi^0$ 
is 16\%. This contribution should be subtracted from the cross section
if it is used to determine the parameters of the 
$\omega(1420)$ and $\omega(1650)$ resonances.

Figure~\ref{spectr} shows that the $\rho(1450)\pi$ contribution 
becomes essential at $\sqrt{s}\approx 1.67$ GeV: $f_{\rho\pi}=(55\pm 6)\%$, 
$f_{\rho^\prime\pi}=(24\pm 7)\%$. The contribution of the interference 
between these states is about $22\%$. The fraction 
$f_{\omega\pi}=(2.1\pm 0.4)\%$, and the interference with isoscalar states
is approximately $-2\%$.

\begin{figure*}
\centering
\resizebox{0.7\textwidth}{!}{\includegraphics{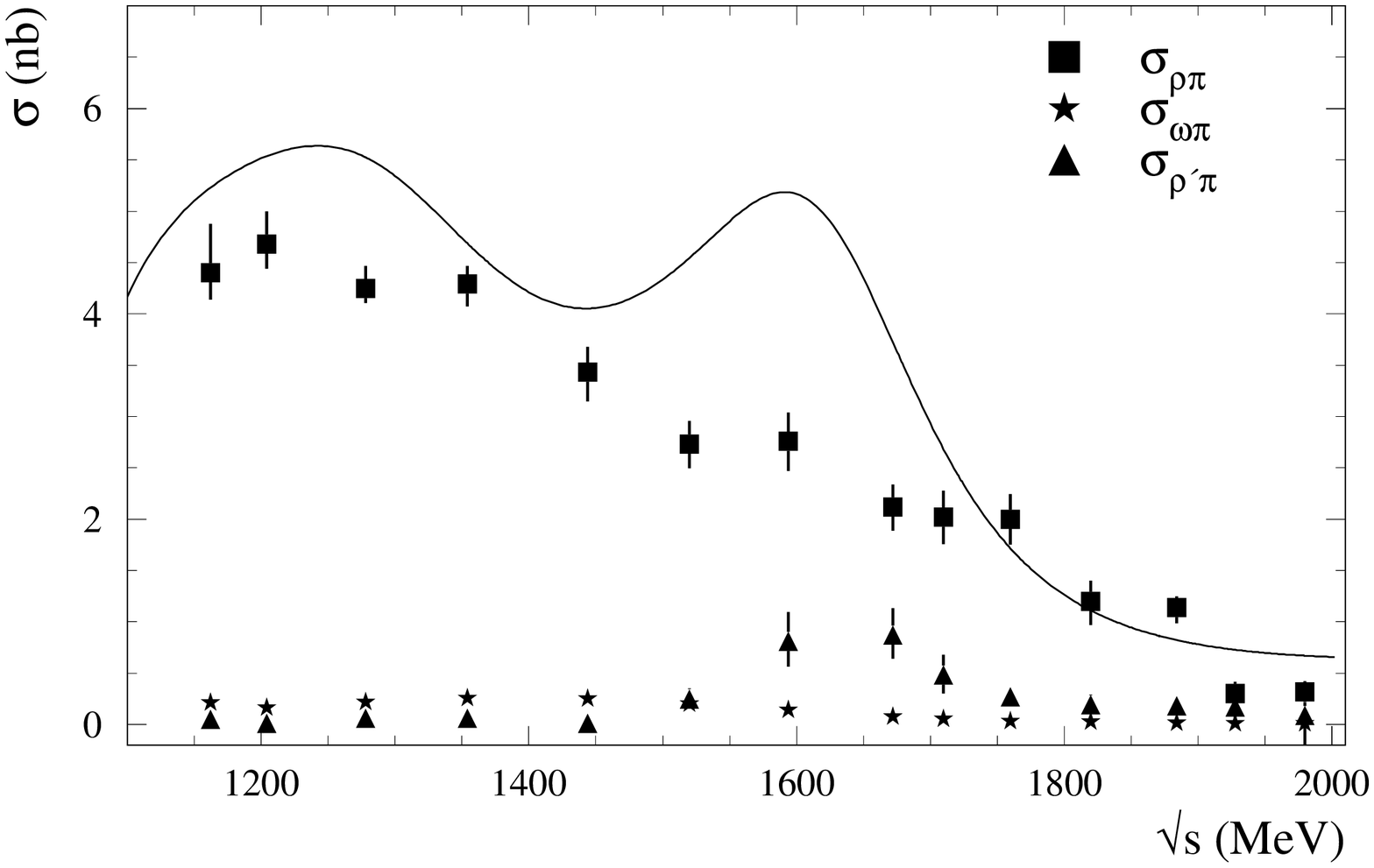}}
\caption{The measured energy dependences of the cross sections
$\sigma_{\rho\pi}$, $\sigma_{\rho^\prime\pi}$, and $\sigma_{\omega\pi}$.
The curve is the result of the fit to the SND data on the total 
$e^+e^-\to\pi^+\pi^-\pi^0$ cross section (see Fig.~\ref{crs-res}).
\label{fit-a1}}
\end{figure*}
The fit parameters for all 14 energy intervals are listed in
Table~\ref{tab2}. In the intervals, in which $\sigma_{\rho^\prime\pi}$ is
consistent with zero, the phase $\phi_1$ cannot be determined from the fit.
Figure~\ref{fit-a1} shows the energy dependences of the cross sections 
$\sigma_{\rho\pi}$, $\sigma_{\rho^\prime\pi}$, and $\sigma_{\omega\pi}$ 
in comparison with the total $e^+e^-\to\pi^+\pi^-\pi^0$ cross section
(curve from Fig.~\ref{crs-res}). It is seen that the cross section 
$\sigma_{\rho^\prime\pi}$ differs from zero in the region of the second
maximum in the $e^+e^-\to\pi^+\pi^-\pi^0$ cross section, corresponding to 
the  $\omega(1650)$ resonance. In the cross section $\sigma_{\rho\pi}$
the resonance structure near 1650 MeV is not seen. We conclude that the 
intermediate state $\rho(1450)\pi$ gives a significant contribution to 
the decay of $\omega(1650)\to\pi^+\pi^-\pi^0$, while the $\rho(770)\pi$
dominates in the $\omega(1420)\to\pi^+\pi^-\pi^0$ decay.

\begin{figure*}
\centering
\resizebox{0.7\textwidth}{!}{\includegraphics{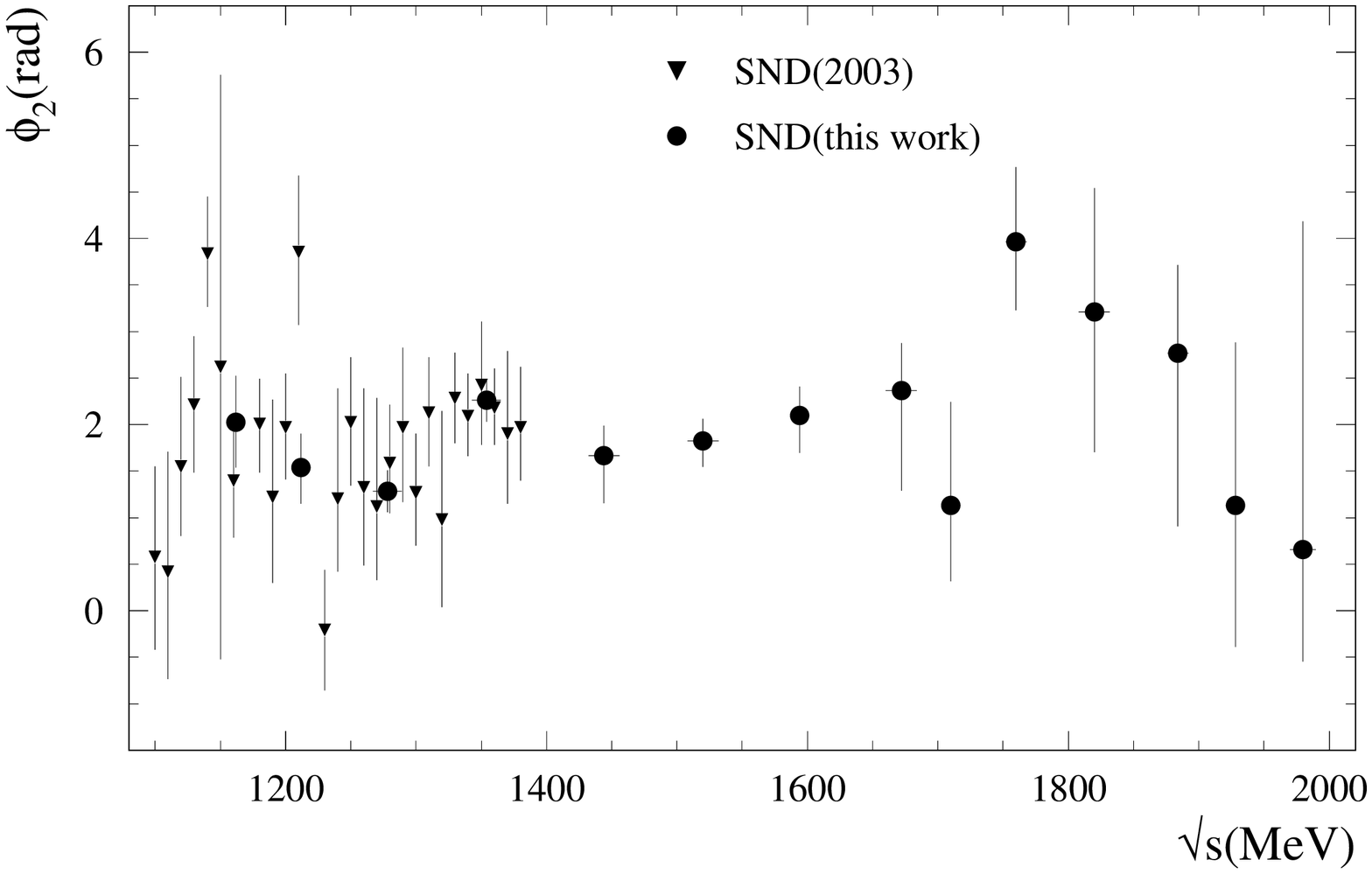}}
\caption{The relative phase between the $\omega\pi^0$ and $\rho(770)\pi$
amplitudes measured in this work in comparison with the results of 
Ref.~\cite{snd2}.
\label{phase_comp}}
\end{figure*}
Figure~\ref{phase_comp} shows the energy dependence of the relative
phase $\phi_2$ between the $\omega\pi^0$ and $\rho(770)\pi$ amplitudes.
In the region $\sqrt{s}=1.15$--1.55 GeV, it is close to $\pi/2$. 
It should be noted that a phase shift of approximately $\pi/2$ is
generated by the $\rho-\omega$ mixing~\cite{tomp3pi-1,snd2}, 
which is the dominant mechanism of the $\omega\to\pi^+\pi^-$ decay.
Below 1.4 GeV our result agrees with the measurement of 
Ref.~\cite{snd2}. In this work, $A_{\omega\pi}$ is parametrized taking
into account the $\rho-\omega$ mixing. For comparison
with our measurement, $\pi/2$ is added to the results from Ref.~\cite{snd2}.

\section{Summary}
In the experiment with the SND detector at the VEPP-2000 $e^+e^-$ collider, the
dynamics of the process $e^+e^-\to\pi^+\pi^-\pi^0$ has been studied in the
c.m. energy range from 1.15 to 2.00 GeV. The $\pi^+\pi^-$ invariant mass 
spectra and the two-dimensional distribution of the momenta of charged pions
have been fitted with the model including the $\rho(770)\pi$, $\rho(1450)\pi$,
and $\omega\pi^0$ intermediate states. The modulus squared of the
$\omega\pi^0$ amplitude has been fixed from our measurement of the 
$e^+e^-\to\omega\pi^0\to \pi^0\pi^0\gamma$ cross section~\cite{SNDompi}.
As a result of the fit, the cross sections for the intermediate states 
$\rho(770)\pi$ and $\rho(1450)\pi$, and the relative phases between 
the $\rho(770)\pi$ amplitude and the $\rho(1450)\pi$ and $\omega\pi^0$
amplitudes have been obtained for 14 energy intervals. The cross section for 
the intermediate state $\rho(1450)\pi$ differs significantly from zero in 
the range 1.55--1.75 GeV, where the resonance $\omega(1650)$ is located. 
In the $\rho(770)\pi$ cross section the resonance structure near 1650 MeV 
is not observed. We conclude that the intermediate state $\rho(1450)\pi$ 
gives a significant contribution to the decay $\omega(1650)\to\pi^+\pi^-\pi^0$,
and that the $\omega(1420)\to\pi^+\pi^-\pi^0$ decay is dominated by
the  $\rho(770)\pi$ intermediate state.

As a result of the refinement of the model describing the 
$e^+e^-\to\pi^+\pi^-\pi^0$ internal structure, the correction has been
determined for
the detection efficiency, which was previously calculated in the model 
of Ref.~\cite{Czyz}. This correction is maximal at $\sqrt{s}=1.8$ GeV,
where it is about 7\%. With this correction the measurement of the
$e^+e^-\to\pi^+\pi^-\pi^0$ cross section based on the 2011 data
set~\cite{snd4} has been updated. The cross section has been also measured
using the 2012 data set. Both measurements are consistent with each
other and with the result of the BABAR experiment~\cite{BABAR}.
The data on the cross section for the process  $e^+e^-\to\pi^+\pi^-\pi^0$, 
obtained in this work, refine and replace the data of Ref.~\cite{snd4}.

\acknowledgement{
This work is supported by the RFBR grant No. 20-02-00060-a.}

\end{document}